\pdfoutput=1
\documentclass[11pt]{article}
\usepackage{amssymb}
\parskip \baselineskip
\usepackage{graphicx}
\newcommand{\ds}{\displaystyle}
\newcommand{\R}{\mathbb{R}}

\newcommand{\ol}{\overline}

\begin{document}

\title{An efficient data structure for counting all linear extensions of a poset, calculating its jump number, and the likes}

\author{Marcel Wild}

\maketitle

\begin{quote}
A{\scriptsize BSTRACT}: {\footnotesize Achieving the goals in the title (and others) relies on a cardinality-wise scanning of the ideals of the poset. Specifically, the relevant numbers attached to the $k+1$ elment ideals are inferred from the corresponding numbers of the $k$-element (order) ideals. Crucial in all of this is a compressed representation (using wildcards) of the ideal lattice. The whole scheme invites distributed computation.}
\end{quote}

\section{Introduction}
The uses of (exactly) counting all linear extensions of a poset are well documented, see e.g. [1] and [9, Part II]. Hence we won't dwell on this, nor on the many applications of the other tasks tackled in this article. Since all of them are either NP-hard or \#P-hard [1], we may be forgiven for not stating a formal Theorem; the numerical evidence of efficiency must do (Section 8). 

It is well known [5, Prop. 3.5.2] that a linear extension corresponds to a path from the bottom to the top of the ideal lattice $Id(P)$ of $P$. Attempts to exploit $Id(P)$ for calculating the number $e(P)$ of linear extensions of $P$ were independently made in [6], [8], [9]. In the author's opinion they suffer from a one-by-one generation of $Id(P)$ and (consequently) an inability or unawareness to retrieve all $k$-element ideals fast. 

Section 2 offers as a cure the compressed representation of $Id(P)$ introduced\footnote{We mention that apart from $Id(P)$ other set systems can be compressed in similar ways. See [12] for a survey.} in [11]. It is based on wildcards and ``multivalued'' rows, as opposed to $01$-rows $=$ bitstrings. This {\it key data structure} is not affected by the sheer size of $Id(P)$ but rather by the number of  arising multivalued rows. As case in point, if $P$ is a 100-element antichain then $|Id(P)| = 2^{100}$ but $Id(P)$ can be represented by the single multivalued row $(2,2, \cdots, 2)$. 

Sections 3 to 6 are dedicated to calculating, respectively, the number $e(P)$ of all linear extensions, the average ranks, the rank probabilities (touching upon the  $\frac{1}{3} - \frac{2}{3}$ conjecture), and the weighted jump number $j(P)$. Section 7 glimpses at further potential uses and generalizations of the key data structure.

\section{The key data structure}
Consider the poset $P_0 = (P_0, \leq)$ in Figure 1. As mentioned, the crucial ingredient of our method is a compressed representation of $Id(P_0)$; this is provided in Table 1.

\begin{center}
\includegraphics{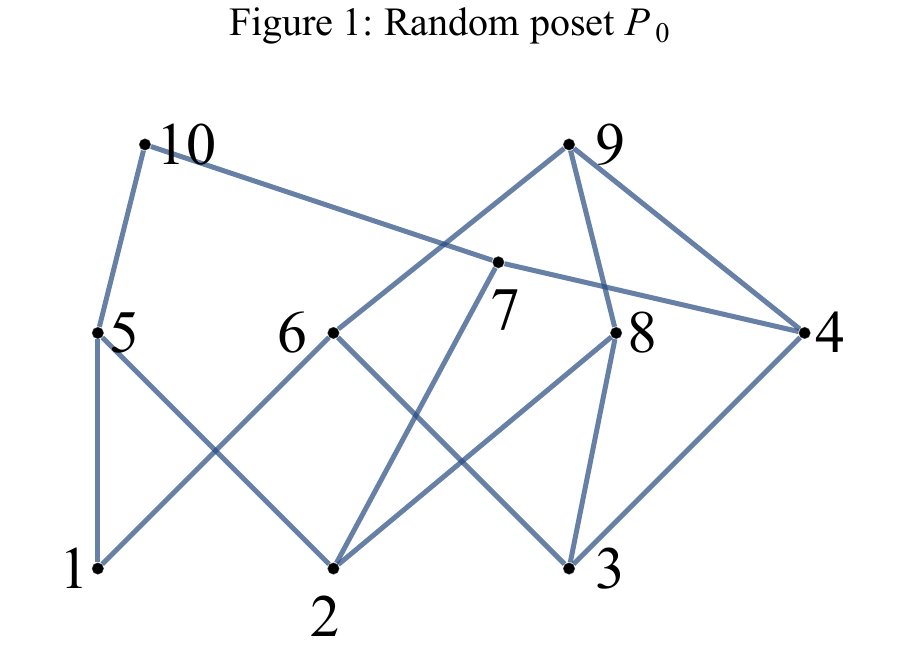}
\end{center}

\begin{tabular}{l|c|c|c|c|c|c|c|c|c|c|}
 & 1 & 2 & 3 & 4 & 5& 6 & 7 &8 & 9 & 10 \\ \hline
 $r_1 =$ & $b_2$ & $b_2$ & $b_1$ & $a_1$ & $a_2$ &0 & 0 &0 & 0 & 0 \\ \hline
 $r_2 =$ & $b$ & 1 & 1 & 2 & $a$ & 0 & 0 & 1 & 0 & 0 \\ \hline
$r_3 =$ & $b$ & 1 & 1 & 1 & $a$ & 0 & 1 & 2 & 0 & 0 \\ \hline
$r_4 =$ & 1 & 1 & 1 & 1 & 1 & 0 & 1 & 2 & 0 & 1 \\ \hline
 $r_5 =$ & 1 & $b$ & 1 & 2 & $a$ & 1 & 0 & 0 & 0 & 0 \\ \hline
 $r_6 =$ & 1 & 1 & 1 & $b$ & 2 & 1 & 0 & 1 & $a$ & 0 \\ \hline
 $r_7 =$ & 1 & 1 & 1 & 1 & $b_1$ & 1 & 1 & $b_2$ & $a_2$ & $a_1$ \\ \hline
  \end{tabular}

Table 1: Compression of $Id(P_0)$ with wildcards

Each subset of $[10] : = \{1, 2, \cdots, 10\}$ will be identified with a length $10$ bitstring in the usual way; for instance $\{ 2, 7, 8\}$ ``$=$'' $(0,1,0,0,0,0,1,1,0,0)$. Each {\it multivalued row} $r = r_i$ represents a certain set of ideals (i.e. their matching bitstrings). Namely, $r$ is the set of all bitstrings $x \in \{0,1\}^{10}$ subject to two types of conditions; either {\it local} or {\it spread-out}. As to local, if the $j$-th entry of $r$ is $0$ or $1$ then $x_j$ must be $0$ or $1$ accordingly. If the $j$-th entry is $2$ then there is no\footnote{Our symbol ``2'' corresponds to the common don't-care symbol ``$\ast$'' but better provides the idea that $x_j$ has {\it two} options.} restriction on $x_j$. As to spread-out, suppose $r$ contains the wildcard $abb \cdots b$ (in any order). That signifies that when the bit of $x$, that occupies the position of $a$, happens to be 1 then also the bits of $x$ occupying the positions of the $b$'s must be 1. A multivalued row can contain several such wildcards, in which case they are independent from each other (and distinguished by subscripts in Table 1). Thus $(1,1,1,1,0,0,0,1,0,0) \in r_2$ but $({\bf 0}, 1,1, 0, {\bf 1}, 0,0,1,0,0) \not\in r_2$ since the $ab$-wildcard is violated. It is straightforward to calculate the cardinality $N : = |Id (P_0)|$ from Table 1:

(1) \quad $N = |r_1| + \cdots + |r_7| = 15 + 6 + 6+ 2 + 6 + 6+ 9 = 50$

What is more, the numbers $N_k$ of $k$-element ideals are readily obtained as coefficients of fairly obvious polynomials. For instance the polynomial for $r_6$ is $x^5(1+x)(1+x+x^2) =x^5 + 2x^6 + 2x^7 + x^8$, and so $r_6$ contains (e.g) two $7$-element, one $8$-element and no 4-element ideal. See [11] for details. Instead of ``key data structure'' we henceforth use the catchier term {\it ideal-coal-mine}. It evokes the picture of an ``ideal-tree'' [2,7] having undergone intense compression.

\section{Counting all linear extensions}

In order to calculate the number $e(P_0)$ of linear extensions of $(P_0, \leq)$ in Figure 1 we  proceed cardinality-wise, but now (as opposed to calculating the numbers $N_k$) need to access all $k$-ideals {\it individually}. Nevertheless the ideal-coal-mine in Table 1 continues to be useful.

First, for each $X \in Id(P_0)$ (or any other $Id(P)$) and any $a \in X$ it can be decided fast whether or not $X \setminus \{a\}$ remains an ideal (and hence a lower cover of $X$ in the lattice $Id(P_0)$). Namely, if $uc(a)$ is the set of all upper covers of $a$ in $(P_0,\leq)$ then clearly

(2) \quad $X \setminus \{a\}$  is an ideal iff $X \cap uc(a) = \emptyset$.

For instance $X = \{1,2, 3, 4, 8\} \in Id(P_0)$ but $X \setminus \{2\} \not\in Id(P_0)$ because $X \cap uc(2) = \{8\} \neq \emptyset$. On the other hand $X \setminus \{1\} \in Id(P_0)$ since $X \cap uc(1) = \emptyset$.

Say that by induction for $k =4$ we obtained the lexicographically ordered list (3) of all pairs $(X, e (X))$ where $X$ ranges over all $k$-element ideals of $P_0$, and $e(X)$ denotes the number of linear extensions of the induced subposet $(X, \leq)$.

(3) \quad $(\{2,3,4,8\}, 5), \ (\{2,3,4,7\}, 3), \ (\{1,3,4,6\}, 5)$, \ $(\{1,2,3,8\}, 8)$, \\
\hspace*{.9cm}  $(\{1,2,3,6\}, 8), \ (\{1,2,3,5\}, 8), \ (\{1,2,3,4\}, 12)$

Here ``lexicographic'' refers to the first components of the pairs: The sets $\{2,3,4,8\}, \{2,3,4,7\}, \cdots,$ when translated to length ten dyadic numbers in the obvious way satisfy $0111000100 (= 452) < 0111001000 (= 456) < \cdots$. The corresponding list for $k+1 = 5$ is obtained by scanning in each row $r_i$ of Table 1 the ideals of cardinality $k+1$. Specifically, suppose that by processing $r_1$ to $r_3$ the list has grown (lexicographically) so far:

(4) \quad $(\{2,3,4,7,8\}, 8), (\{1,2,3,5,8\}, 16), (\{1,2,3,4,8\}, 25), (\{1,2,3,4,7\}, 15), (\{1,2,3,4,5\}, 20)$

We turn to $r_4$ and notice that it doesn't contain cardinality $k+1$ ideals. But $r_5$ has two of them, i.e. $X = \{1,2,3,5,6\}$ and $Y = \{1,2,3,4,6\}$. Evidently

(5) \quad $e(X) = e(X_1) + e(X_2) + \cdots$

where $X_1, X_2,\cdots$ are the lower covers of $X$ in $Id(P_0)$. The lower covers of $X$ are among the sets $X \setminus \{a\} \ (a \in X)$. Applying criterion (2) one checks (see Figure 1) that only $X_1 = X \setminus \{5\} = \{1,2,3,6\}$ and $X_2 = X \setminus \{6\} = \{1,2,3,5\}$ qualify. To find out the values $e(X_1)$ and $e(X_2)$ needed in (5) we look for the paris $(X_1, ?)$ and $(X_2, ?)$ in list (3). Generally speaking, using binary search to locate an element in an ordered list of length $\ell$ takes time $O(\ell og(\ell))$. We obtain that $e(X) = e(X_1) + e(X_2) = 8 +8 = 16$. Similarly $Y$ has the lower covers 
$$Y_1 = Y\setminus \{2\} = \{1, 3, 4, 6\}, \ \ Y_2 = Y \setminus \{4\} = \{1, 2, 3, 6\}, \ \ Y_3 = Y \setminus \{6\} = \{1,2, 3, 4\}$$
and after consulting (3) one obtains $e(Y) = e(Y_1) + e(Y_2) + e(Y_3) = 5 + 8 + 12 = 25$. The pairs $(X, 16)$ and $(Y, 25)$ are now inserted in list (4) at the right place. After processing rows $r_6$ and $r_7$ in Table 1 the same way as $r_5$ one obtains the analogon of list (3) for $k+1$. List (3) can now be discarded. In the end the list for $k =|P_0| = 10$ contains just one pair, i.e. $(P_0, e(P_0)) = (P_0, 2212)$.

{\bf 3.1} As to distributed computation  ($=$ parallelization), suppose the sorted list ${\cal L}$ of pairs $(X, e(X))$, where $X$ ranges over the $k$-element ideals, has been compiled at the control unit of a distributed network. The control unit then sends ${\cal L}$ to all satellites $S_1, S_2, \cdots$ and (according\footnote{The amount of work for each satellite can be predicted accurately since as seen in Section 2 the number of $k+1$-element ideals in any multivalued row is easily calculated.} to their capacities) distributes the rows of the ideal-coal-mine among them. Each satellite $S_i$ sieves the $(k+1)$-element ideals $Y$ from its rows and, by using ${\cal L}$ and (5), creates a sorted list ${\cal L}_i$ of pairs $(Y, e(Y))$. The lists ${\cal L}_1,{\cal L}_2 \cdots$ are sent back to the control unit where they are merged to a sorted list ${\cal L}'$. This completes the cycle. In the same way {\it all} tasks to be discussed in this article are susceptible to distributed computation.



\section{Calculating the average ranks}

Given an ideal $X$ of $P$ and a linear extension ${\cal E}$ of the induced poset $(X, \leq)$, we define the {\it rank} $r (X, {\cal E}, a)$ of any $a \in X$ (w.r.to $X$ and ${\cal E}$) as the position that $a$ occupies in ${\cal E}$. The {\it average rank} of $a$ in $X$ is defined as

(6) \quad $avr(X,a) : = \ds\frac{1}{e(X)} \sum \{r(X, {\cal E}, a): \ {\cal E} \in LE(X)\}$ \quad (see e.g. [1])

where $LE(X)$ denotes the set of all linear extensions ${\cal E}$ of $(X,\leq)$. Thus $|LE(X)| = e(X)$.
In order to calculate $avr(X, a)$ recursively we let $c_1, \cdots, c_t$ be the maximal elements of $(X, \leq)$ and make a case distinction.

{\it Case 1:} $a \not\in \{c_1, \cdots, c_t\}$.
Putting $Y_i : = X\setminus \{c_i\}$ as well as $\gamma_i : = e(Y_i) / e(X)$ for $1 \leq i \leq t$, we claim that

(7) \quad $avr(X, a) = \gamma_1 \, avr(Y_1, a) + \cdots + \gamma_t \, avr (Y_t, a)$

{\it Proof of (7)}. The set $LE(X)$ gets partitioned into $t$ parts according to which element $c_i$ appears last in ${\cal E}$. Since $a$ is not maximal we have $a \in Y_i$ for all $1 \leq i \leq t$. Hence if ${\cal E} \in LE(X)$ is of type ${\cal E} = (\cdots, c_i)$ and if ${\cal E}| Y_i $ by definition results by dropping $c_i$ from ${\cal E}$, then   ${\cal E}| Y_i \in LE(Y_i)$ and $r(X, {\cal E}, a)  = r(Y_i, {\cal E} | Y_i, a)$. Conversely each ${\cal E}' \in LE(Y_i)$ is of type ${\cal E}' = {\cal E} |Y_i$. Consequently

$$\begin{array}{lll}
e(X)\cdot avr(X,a) &  {(6) \atop =} & \sum \{r(X, {\cal E}, a): {\cal E} \in LE(X) \} \\ \\
 &= & \sum \{r(Y_1, {\cal E}', a): \ {\cal E}' \in LE(Y_1) \} + \cdots + \sum \{r(Y_t, {\cal E}', a) : \ {\cal E}' \in LE(Y_t) \} \\
 \\
  & {(6) \atop =} & e(Y_1) avr (Y_1, a) + \cdots + e(Y_t) avr (Y_t, a). 
\end{array}$$
The claim follows upon division with $e(X)$ throughout. \quad $\square$

{\it Case 2:} $a \in \{c_1, \cdots, c_t\}$, say without loss of generality $a = c_1$. We claim that

(8) \quad $avr (X, a) \  = \ \gamma_1 \ |X| + \gamma_2 \, avr (Y_2, a) + \cdots + \gamma_t \, avr (Y_t, a)$

{\it Proof of (8)}. We consider the same partitioning of $LE(X)$ as in the proof of (7). Yet now there are ${\cal E} \in LE(X)$ that {\it end} in $a$. As for any $c_i$ their number is $e(X \setminus \{a\}) = e(X \setminus \{c_1\}) = e(Y_1)$ but evidently $r(X, {\cal E}, a) = |X|$ for each such ${\cal E}$. Consequently

$e(X) avr(X, a) \ {(6) \atop =} \ \sum (r(X, {\cal E}, a): \ {\cal E} \in LE(X) \}$
 
$ \ = \ |X| e((Y_1) + \sum (r(Y_2, {\cal E}', a) : \ {\cal E}' \in LE(Y_2) \} + \cdots + 
 \sum \{r(Y_t, {\cal E}', a): \ {\cal E}' \in LE(Y_t) \}$
 
$ \ {(6) \atop = } \  |X| e(Y_1) + e(Y_2) avr (Y_2, a) + \cdots + e(Y_t) avr (Y_t, a)$

 from which the claim follows upon dividing by $e (X)$. \quad $\square$

\section{Rank probabilities and the $\frac{1}{3} - \frac{2}{3}$ conjecture}

Let $e(P | a, k)$ be the number of linear extensions ${\cal E}$ of $P$ in which $a$ occupies the $k$-th position. The parameters $e(P| a, k)$ can be calculated recursively akin to Section 4. Hence $p_a(k): = e(P| a, k) / e(P)$ is the ({\it absolute}) {\it rank probability}  that $a$ occupies the $k$-th rank in a random linear extension.  Basic probability theory yields

(9) \quad $avr (P, a) = \ds\sum_{k=1}^n k p_a (k) \quad (n : = |P|)$.

Equation (9) thus yields the average ranks as a side product of the $n^2$ rank probabilities. The extra information provided by these probabilities may however not justify the effort computing them when $n$ gets large.

{\bf 5.1} In the remainder of Section 5 we calculate the {\it relative rank probability} $p(a,b)$ that $a$ precedes $b$ in a random linear extension ${\cal E}$. Since $p(a,b) = 1$ when $a < b$ in $(P, \leq)$, and $p(a,b) =0$ when $a > b$ in $(P, \leq)$, we henceforth focus on incomparable $a, b \in P$. Obviously $p(a,b) = e(P, a, b)/e(P)$ where generally for any ideal $X \subseteq P$ we define $e(X, a, b)$ as the number of linear extensions of $(X, \leq )$ where $a$ precedes $b$. In order to calculate $e(X, a, b)$ recursively we  let $c_1, \cdots, c_t$ be the maximal elements of $X$ and put $Y_i : = X \setminus \{c_i\}$ for all $1 \leq i \leq t$.

{\it Case 1:} Neither $a$ nor $b$ are maximal in $X$. Then $a, b \in Y_i$ for all $1 \leq i \leq t$, and obviously

(10) \quad $e(X, a, b) = e(Y_1, a, b) + \cdots + e(Y_t, a, b)$

{\it Case 2:} $a$ is maximal (say $a = c_1$) but not $b$. Then in {\it none} of the $e(Y_1)$ many linear extensions of $(X, \leq)$ that end in $a$, we have $a$ preceeding $b$, and so

(11) \quad $e(X, a, b) = e(Y_2, a, b) + \cdots + e(Y_t, a, b)$

{\it Case 3:} $b$ is maximal (say $b = c_1)$ but not $a$. Then in {\it all} $e(Y_1)$ many linear extensions of $(X, \leq)$ that end in $b$, we have $a$ preceeding $b$, and so

(12) \quad $e(X, a,b) = e(Y_1) + e(Y_2, a, b)  + \cdots + e(Y_t, a, b)$

{\it Case 4:} Both $a$ and $b$ are maximal. If say $a=c_1$ and $b = c_2$ then 

(13) \quad $e(X, a,b) = e(Y_2) + e(Y_3, a, b) + \cdots + e(Y_t, a, b)$

{\bf 5.2} The famous $\frac{1}{3} - \frac{2}{3}$ conjecture states that for every poset $(P, \leq)$ which is not linearly ordered, there are elements $a, b \in P$ such that $\frac{1}{3} < p (a,b) < \frac{2}{3}$ (and whence also $\frac{1}{3} < p(b,a) < \frac{2}{3}$). If this conjecture is false (it fails for infinite posets) then our fast algorithm for calculating all probabilities $p(a,b)$ might be helpful in finding a counterexample.

\section{Calculating the weighted jump number}

Recall that in any linear extension ${\cal E} = (a_1, a_2, \cdots, a_n)$ of a $n$-element poset $P$ the pair $(a_i, a_{i+1})$ is called a {\it jump} if $a_{i+1}$ is {\it not} an upper cover of $a_i$. Suppose that associated with each ordered pair $(a,b)$ of incomparable elements of $P$ is a penalty $pen(a,b) \in \R^+$. Define $j({\cal E})$ as the sum of all numbers $pen(a_i, a_{i+1})$ where $(a_i, a_{i+1})$ ranges over the jumps of ${\cal E}$. Further call $j(P) : = \min \{ j ({\cal E}) : {\cal E} \ \mbox{is linear extension of} \ P\}$ the {\it weighted jump number} of $(P, \leq)$. If all $pen(a,b)$ are set to 1 then $j(P)$ is the ``ordinary'' jump number of $P$, i.e. the minimum number of jumps occuring in any linear extension of $P$. 

{\bf 6.1} Let us see how our framework for calculating $e(P)$ caters for $j(P)$ as well. Consider some $m$-element ideal $X$ of $P$ and some linear extension ${\cal E} = (a_1, \cdots, a_m)$ of $(X, \leq)$. Then $a_m$ is a maximal element of $X$ and ${\cal E}'= (a_1, \cdots, a_{m-1})$ is a linear extension of the ideal $X \setminus \{a_m\}$. If $a_{m-1}$ is a lower cover of $a_m$  then $j({\cal E}) = j({\cal E}')$. Otherwise $j({\cal E}) =j({\cal E}') + pen(a_{m-1}, a_m)$. 

For each ideal $X$ and each maximal element $b \in X$ let $j(X, b)$ be the minimum of all numbers $j({\cal E})$ where ${\cal E}$ ranges over all linear extensions of $X$ of type ${\cal E} = (\cdots, b)$. If we manage to calculate all $j(X, b)$ recursively then $j(P)$ will be obtained as the minimum of all numbers $j(P, b)$ where $b$ ranges over the maximal elements of $P$.

As to calculating $j(X, b)$,  putting $Y : = X \setminus \{b\}$ we see that $j(X, b)$ can be obtained from $Y$ as follows. Let $c_1, \cdots, c_s$ be the maximal elements of $Y$ which happen to be lower covers of $b$ (possibly there are none), and let $c_{s+1}, \cdots, c_t$ be the remaining maximal elements of $Y$ (possibly there are none). Then

(14) \quad $j(X, b) = \min \{j (Y, c_1), \cdots, j(Y, c_s), \ j (Y, c_{s+1}) + pen(c_{s+1}, b), \cdots, j(Y, c_t) + pen(c_t, b)\}$

The formally best algorithm [10] for calculating $j(P)$ has complexity $O(1.8638^n)$ and apparently has not yet been implemented.  In fact the only implemented and published algorithm prior to the present article seems to be [4]. It uses so-called greedy chains of $(P, \leq)$ to calculate the ordinary jump number.  A generalization of [4] to the weighted case is not straightforward.

{\bf 6.2} Apart from calculating the number $j(P)$, how can we {\it get} an optimal linear extension ${\cal E}_0$, i.e. satisfying $j({\cal E}_0) =j(P)$? As for general dynamic programming tasks, proceed as follows. After each type (14) update store the element $c_i$ that achieves $j(X, b) =j(Y, c_i)$, respectively $j(X,b) = j(Y, c_i) + pen(c_i, b)$. Ties are broken arbitrarily. After the so enhanced algorithm of 6.1 has finished, do the following. Starting with $X = P$, and always picking the lower cover of $X$ determined by the pinpointed element, yields ${\cal E}_0$.

\section{Further applications and generalizations}

We glimpse at two further applications of the ideal-coal-mine: Scheduling with time-window constraints (7.1), and the risk polynomial of a poset (7.2). We also speculate on generalizing all tasks discussed in this article from posets to antimatroids (7.3).

{\bf 7.1} Instead of penalities suppose that coupled to each $a \in P$ is a positive number $T(a)$ which can be interpreted as the duration to complete job $a$. For each $Y \subseteq P$ put $T(Y) = \Sigma \{(T(a): a \in Y\}$. Suppose that each job $a$ needs to be completed within a time-window $W(a)$. For each ideal $X$ let $e^\ast(X)$ be the number of all linear extensions $(a_1, \cdots, a_m)$ of $(X, \leq)$ that satisfy $T(a_1) + T(a_2) + \cdots + T(a_i) \in W(a_i)$ for all $1 \leq i \leq m$. Evidently $e^\ast (X)$ can again be obtained recursively as the sum of all numbers $e^\ast(X \setminus \{a\})$ where $a \in \max (X)$ is such that $T(X \setminus \{a\}) + T(a)$ happens to be in $W(a)$. 

In a similar vein, but more involved, one can calculate the (weighted) jump number restricted to the {\it subset} of all linear extensions ${\cal E} \in LE(P)$ that satisfy the time-window constraints. See also [2].

{\bf 7.2} In the nice math-biology mix [7], where e.g. spaces of genotypes are modelled as distributive lattices, a crucial r\^{o}le is placed by the so called {\it risk polynomial} ${\cal R} (P, {\bf f})$ of $(P, \leq)$. (Admittedly the following remarks may be too vague for readers unfamiliar with [7] but they may provide a flavour of things.) The many\footnote{However, for some applications plenty variables are equated.} variables $f_I$ of ${\cal R} (P, {\bf f})$ are indexed by the nontrivial ideals $I$ of $P$. Furthermore, by [7, Thm.15] ${\cal R}(P, {\bf f})$ equals the sum of certain  products ${\bf f}(\pi)$  where $\pi$ runs over $LE(P)$.
The precise definition of ${\bf f}(\pi)$ in [7, eq.(12)], corrobarated by  [6, Example 16], seems to indicate that scanning $LE(P)$ can be avoided by processing the much fewer {\it filters} $F$ of $(P, \leq)$. Namely (dual to what we did with ideals), considering $F$ as a poset $(F, \leq)$ and letting $e_1, \cdots, e_t$ be its minimal elements, the risk polynomial ${\cal R}(F, {\bf f})$ (whose variables are thus indexed by ideals of $(F, \leq)$) accordingly decomposes as a sum of polynomials
$${\cal R}(F, {\bf f}) = {\cal R}(F, {\bf f}, e_1) + \cdots + {\cal R}(F, {\bf f}, e_t).$$
If $F_1$ is the smaller filter $F \setminus \{e_1\}$ (similarly $F_i$ for $i \geq 2$ is defined), say with minimal elements $\ol{e}_1, \ol{e}_2, \cdots$, then ${\cal R}(F, {\bf f}, e_1)$ arises from the polynomials ${\cal R}(F_1, {\bf f}, \ol{e}_1), {\cal R}(F_1, {\bf f}, \ol{e}_2), \cdots$ in natural ways.

{\bf 7.3} The ideal lattice $Id(P)$ of a poset $(P, \leq)$ is an example of a set system ${\cal A} \subseteq {\cal P}(S)$ which is union-closed and {\it graded} in the sense that all maximal chains from $\emptyset$ to $S$ have the same length $|S|$. Such set systems are precisely the set systems of all feasible sets of an antimatroid. Antimatroids are important structures in combinatorial optimization.  The so called {\it basic words} are to antimatroids what linear extensions are to posets. Chances are good that the ideal-coal-mine carries over from the poset level to the antimatroid level.

\section{Numerics}

The author coded both the ideal-coal-mine (Section 2) and the count of linear extensions (Section 3) as Mathematica 11.0 notebooks\footnote{Using an Intel i5-3470 CPU processor with 3.2 GHz.}. For instance, let us look at a randomly generated ``thin'' 180-element poset $(P, \leq)$, i.e. consisting of 45 levels, each of cardinality 4, such that each element in level $i$ (except for $i=1$) has exactly two lower covers in level $i-1$. It took 9 seconds to display the $3396086$ ideals in  $6970$ multivalued rows. Processing the ideals one-by-one took $18122$ seconds and  yielded

(15) \quad $e(P_1) = 59 \cdots 788 800 \approx 10^{94.8}$

 We chose a thin poset in order to have many small $N'_k$'s, as opposed to few large ones. In particular the highest $N_k$ was $N_{147} = 80'134$. The effect is that the type (3) and (4) lists don't get too long\footnote{As mentioned in 3.1, the type (4) lists can be made as short as pleased by parallelizing. Not so the type (3) lists, but they are only subject to binary search (as opposed to binary search {\it and} insertion).}.  That the magnitude of the $N_k$'s is important becomes apparent in the next example where $(P_2, \leq)$ is obtained by cutting top and bottom of the $64$-element Boolean lattice. We calculated $|ID(P_2)| = 7828 352$, i.e. the sixth Dedekind number, in 16 seconds (using 24871 multivalued rows).
 
 The largest level of Id$(P_2)$ is the middle one with $N_{31} = 492288$ and level $20$ to level $42$ all have cardinality $\geq 100'000$. Although $7828352$ is (roughly) only 2.3 times $3396 086$ it took 3.2 times longer than for $e(P_1)$ to calculate
 
 (16) \quad $e(P_2) = 141 377 \cdots 480 \approx 10^{53.2}$
 
 This number confirms the number reported in [9, p.129], whose computation on a computer server took 16 hours, thus about our time. We mention that Neil Sloane's ``Integer Sequences'' website features $e(P_2)$, as well as plenty other numbers $e(P)$ of lesser interest. Let us only best one of them. Taking as $P_3$ the first five levels of the Fibonacci 1-differential poset $z(1)$ Kavvadias computed $e(P_3) = 1093025200$. We confirmed this number in 0.5 seconds and went on to calculate $e(P_4 = 272'750'206'765'993'342'848$ in 55 seconds where $P_4$ consists of the bottom six levels of $z(1)$. Also $e(P_5)$ (where the $52$-element poset $P_5$ consists of the bottom seven levels) is within reach because $|Id (P_5) | = 35296517$ (calculated in 1.5 sec) is not outrageously high. Trouble is, as discussed above, that $N_{35} = 3068802$ is too high to be handled without parallelizing.

For an updated version of the present article the author sollicites interesting proposals of jump number computations (or the other parameters discussed).

\section*{References}
\begin{enumerate}
\item[{[1]}] G. Brightwell, P. Winkler,  Counting linear extensions, Order 8 (1991) 225-242.
\item[{[2]}] A. Mingozzi, Bianco, S. Ricciardelli, Dynamic programming strategies for the Traveling Salesman Problem with time window and precedence constraints, Oper. Res. 45 (1997) 365-377.
\item[{[3]}] M. Habib, L. Nourine, Tree structure for distributive lattices and its applications. Theoretical Computer Science 165 (1996) 391-405.
\item [{[4]}] L. Bianco, P. Dell'Olmo, S. Giordani, An optimal algorithm to find the jump number of partially ordered sets, Computational Optimization and Applications 8 (1997) 197-210.
	\item[{[5]}] R.P. Stanley, Enumerative Combinatorics, Volume 1, Cambridge Studies in Advanced Mathematics 49 (1997).
		\item[{[6]}] M. Peczarski, New results in minimum-comparison sorting, Algorithmica 40 (2004) 133-145.

	\item[{[7]}] N. Beerenwinkel, N. Eriksson, B. Sturmfels, Evolution on distributive lattices, Journal of Theoretical Biology 242 (2006) 409-420.

	\item[{[8]}] K. De Loof, B. De Baets, H. De Meyer, Exploiting the lattice of ideals representation of a poset, Fundamenta Informaticae 71 (2006) 309-321.
\item[{[9]}] O. Wienand, Algorithms for symbolic computation and their applications, PHD, University of Kaiserslauten 2011.

	\item[{[10]}] D. Kratsch, S. Kratsch, The jump number problem: Exact and parametrized, Lecture Notes in Computer Science 8246 (2013) 230-242.

	\item[{[11]}] M. Wild, Output-polynomial enumeration of all fixed-cardinality ideals of a poset, respectively all fixed-cardinality subtrees of a tree, Order 31 (2014) 121-135.

	\item[{[12]}] M. Wild, ALLSAT compressed with wildcards, Part 1: Converting CNF's to orthogonal DNF's, preliminary version, available on ResearchGate.
\end{enumerate}
\end{document}